\documentclass[aps,pra,groupedaddress,superscriptaddress,twocolumn]{revtex4-1}

\usepackage[utf8x]{inputenc}
\usepackage{color}
\usepackage{bbm} 

\usepackage{amsfonts,amsmath,amssymb,stmaryrd}

\usepackage{graphicx}
\usepackage{subfigure}  
\usepackage{bbm}
\usepackage{hyperref}
\usepackage{epsfig}
\usepackage{mathrsfs}
\usepackage{verbatim}
\usepackage{ulem}	
\usepackage{siunitx}
\usepackage{soul} 

\usepackage{array}

\begin{document}
\normalem	

\title{Anomalous excitation facilitation in inhomogeneously broadened Rydberg gases}

\author{F. Letscher}
\affiliation{Department of Physics and research center OPTIMAS, University of Kaiserslautern, Germany}
\affiliation{Graduate School Materials Science in Mainz, Gottlieb-Daimler-Strasse 47, 67663 Kaiserslautern, Germany}

\author{O. Thomas}
\affiliation{Department of Physics and research center OPTIMAS, University of Kaiserslautern, Germany}
\affiliation{Graduate School Materials Science in Mainz, Gottlieb-Daimler-Strasse 47, 67663 Kaiserslautern, Germany}

\author{T. Niederprüm}
\affiliation{Department of Physics and research center OPTIMAS, University of Kaiserslautern, Germany}

\author{H. Ott}
\affiliation{Department of Physics and research center OPTIMAS, University of Kaiserslautern, Germany}

\author{M. Fleischhauer}
\affiliation{Department of Physics and research center OPTIMAS, University of Kaiserslautern, Germany}

\begin{abstract}
When atomic gases are laser driven to Rydberg states in an off resonant way, a single Rydberg atom may enhance the excitation rate of surrounding atoms. This leads to a facilitated excitation referred to as Rydberg anti-blockade. In the \textit{usual facilitation} scenario, the detuning of the laser from resonance compensates the interaction shift. Here, we discuss a different excitation mechanism, which we call \textit{anomalous facilitation}. This occurs on the "wrong side" of the resonance and originates from inhomogeneous broadening. The anomalous facilitation may  be seen in experiments of attractively interacting atoms on the blue detuned side, where facilitation is not expected to appear.
\end{abstract}

\date{\today}

\maketitle

\section{Introduction}

Ultracold Rydberg gases are an ideal platform to study many-body physics with strong and long-range interactions. Current measurement techniques allow to observe their excitation dynamics \cite{Urvoy2015, Weber2015, Valado2016, Simonelli2016} as well as their spatial correlations \cite{Schauss2012, Schauss2014}. Recently, the so called anti-blockade regime with off resonant laser driving triggered many interesting experimental \cite{Carr2013, Schempp2014, Malossi2014, Urvoy2015, Valado2016} as well as theoretical \cite{Lesanovsky2014a, Mattioli2015, Sibalic2016} studies. While in the blockade regime \cite{Lukin2001,Gorshkov2011,Dudin2012a,Heidemann2007,Ebert2015,Weber2015}, using a resonant drive, a single Rydberg excitation suppresses further excitations within the so called blockade radius, the anti-blockade regime \cite{Ates2007, Ates2007a, Amthor2010} facilitates excitations at a certain distance. In this case the level shift caused by the Rydberg interaction compensates the detuning of the driving laser. This leads to the fast growth of Rydberg excitation clusters \cite{Schempp2014,Malossi2014,Urvoy2015,Valado2016,Simonelli2016,Letscher2016}. 

In many ultracold Rydberg gas experiments strong excitation line broadening limits the observation of coherent dynamics. In particular, there are strong indications of additional interaction induced broadening mechanisms \cite{ Singer2004, Raitzsch2009, Li2013, Macri2014, Goldschmidt2016,Letscher2016}. Moreover, investigations of thermal Rydberg gases inherently show strongly broadened spectral lines \cite{Baluktsian2013, Urvoy2015}. Although these mechanisms typically prevent coherent dynamics, they may lead to novel interesting features in the incoherent dynamics.

Here, we discuss an anomalous facilitation mechanism of Rydberg excitations due to strong inhomogeneous broadening which leads to an enhanced excitation rate on the "wrong" side of the resonance. While the usual facilitation, where the detuning compensates the interaction shift, was observed in previous experiments \cite{Schempp2014, Malossi2014, Urvoy2015,Letscher2016}, here we show the possibility of an anomalous facilitation. Exemplary, we discuss the case of attractively interacting Rydberg atoms with strong inhomogenoeus broadening and show that an excitation cascade is possible on the blue detuned side of the resonance. The additional inhomogeneous broadening is motivated by a recent experiment \cite{Goldschmidt2016} and the idea of interaction induced broadening. 

\begin{figure}[b]
\centering
\epsfig{file=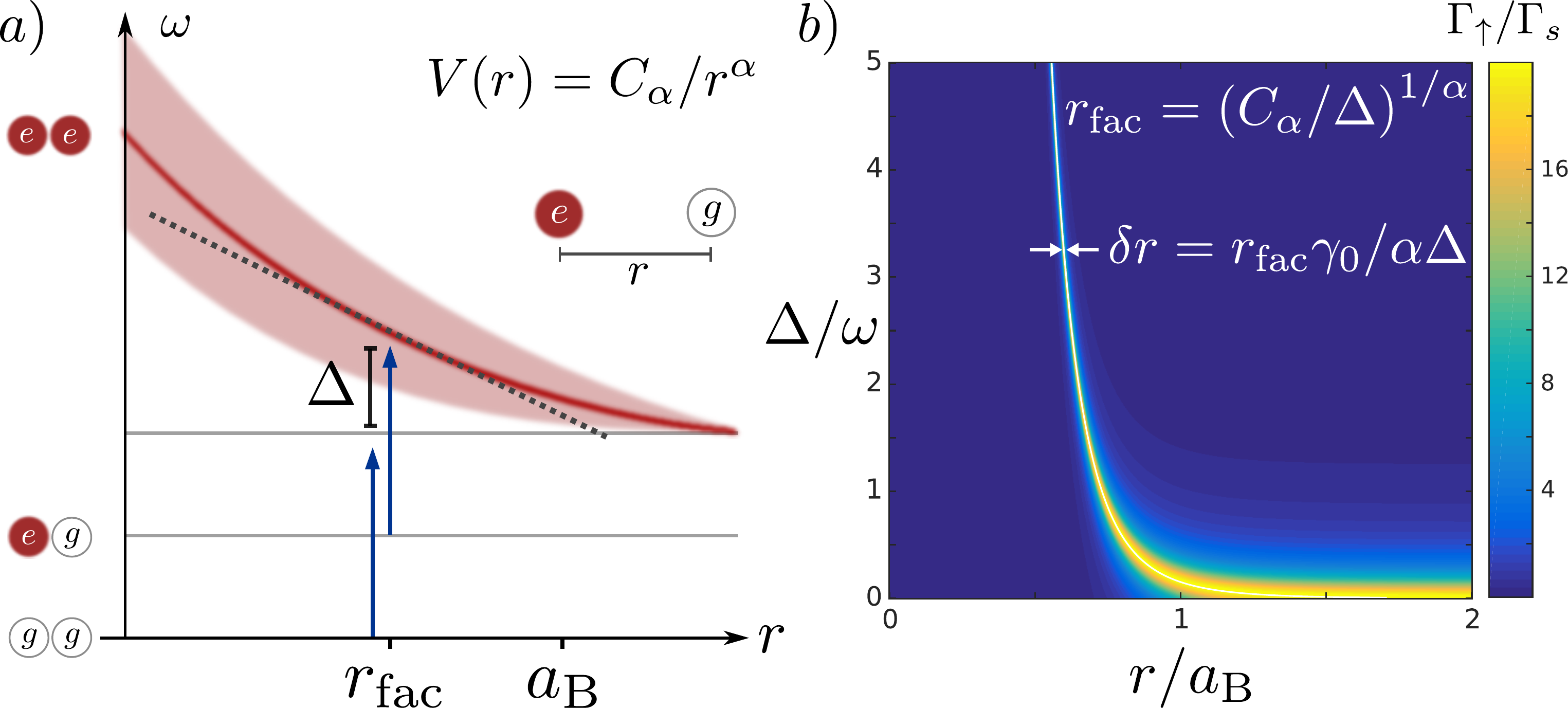, width=.45\textwidth}
\caption{(a) Schematic overview of the two-step excitation mechanism in the incoherent anti-blockade regime. Here, an inhomogeneous broadening is indicated by the red shaded region. At the facilitation radius $r_\mathrm{fac}$, the detuning compensates the interaction shift, i.e. $\Delta = V(r_\mathrm{fac})$. The blockade radius $a_\mathrm{B} = (C_\alpha/\gamma_0)^{(1/\alpha)}$ is always larger than the facilitation radius $r_\mathrm{fac}$. (b) Incoherent excitation rate in the presence of one Rydberg excitation with distance $r$ for different detunings $\Delta$. We set the Rabi frequency $\Omega = 1$ as the natural energy scale and $a=1$ the characteristic length scale. In these units, the parameters are $\Gamma_\mathrm{s} = 0.05$, $\gamma_0 = 2$ and $C_6 = 10\omega$ with linewidth $\omega = \gamma_0 \sqrt{\Omega^2/\Gamma_\mathrm{s}\gamma_0+1}$. }
\label{fig:Fig1}
\end{figure}

\section{Two particle dynamics: usual case}

Let us briefly review the usual facilitation mechanism. Specifically, we discuss two repulsively interacting atoms on the blue detuned ($\Delta > 0$) side of the resonance. To this end, we consider the incoherent excitation rate $\Gamma_\uparrow(r)$ of an atom at distance $r$ from an excited Rydberg atom. It has been shown, that a rate equation approximation successfully describes current experiments \cite{Schempp2014, Weber2015, Urvoy2015}. Due to the long-range interactions, a single Rydberg excitation dramatically alters the excitation rate $\Gamma_\uparrow(r)$ of surrounding atoms. We consider repulsive interactions between Rydberg atoms of the form $V(r) = C_\alpha/r^\alpha > 0$ ($\hbar = 1$) with $\alpha = 6$, corresponding to van-der Waals type interaction. By eliminating all coherences between the ground and Rydberg state and including the interaction effect using an effective detuning $\Delta-V(r)$ \cite{Ates2007a, Hoening2013, Petrosyan2013, Schoenleber2014}, we obtain the excitation rate
\begin{equation}
\label{eq:eq1}
\Gamma_\uparrow(r) = \frac{2\Omega^2\gamma_0}{\gamma_0^2+(\Delta - V(r))^2}.
\end{equation}
The excitation rate $\Gamma_\uparrow(r)$ strongly depends on the distance $r$ between ground state atom and excited Rydberg state. Here, $\Omega$ denotes the Rabi frequency and $\gamma_0$ is the decoherence rate including spontaneous decay and dephasing. 

Fig. \ref{fig:Fig1}a illustrates the incoherent anti-blockade regime starting from both atoms in the ground state. While the first excitation is strongly off resonant, the detuning $\Delta$ can compensate the interaction shift $V(r)$ exciting the second atom to the Rydberg state. For a fixed detuning $\Delta$, the subsequent excitation rate $\Gamma_\uparrow(r)$ reaches its maximum at the so called facilitation radius 
\begin{equation}
r_\mathrm{fac} = (C_\alpha/\Delta)^{(1/\alpha)}.
\end{equation}
Assuming a linear slope, the spatial extent of the facilitation region is given by the expression
\begin{equation}
\delta r \simeq r_\mathrm{fac} \gamma_0/ \alpha\Delta.
\end{equation}
As a consequence the spatial extent $\delta r$ of the facilitation region becomes smaller and smaller if the detuning $\Delta$ is increased. This can be seen in Fig. \ref{fig:Fig1}b, where we show the excitation rate \eqref{eq:eq1} for different detunings $\Delta$ illustrating the transition between blockade regime ($\Delta \lesssim \gamma_0$) to anti-blockade regime ($\Delta \gtrsim \gamma_0$). 

\begin{figure}[b]
\centering
\epsfig{file=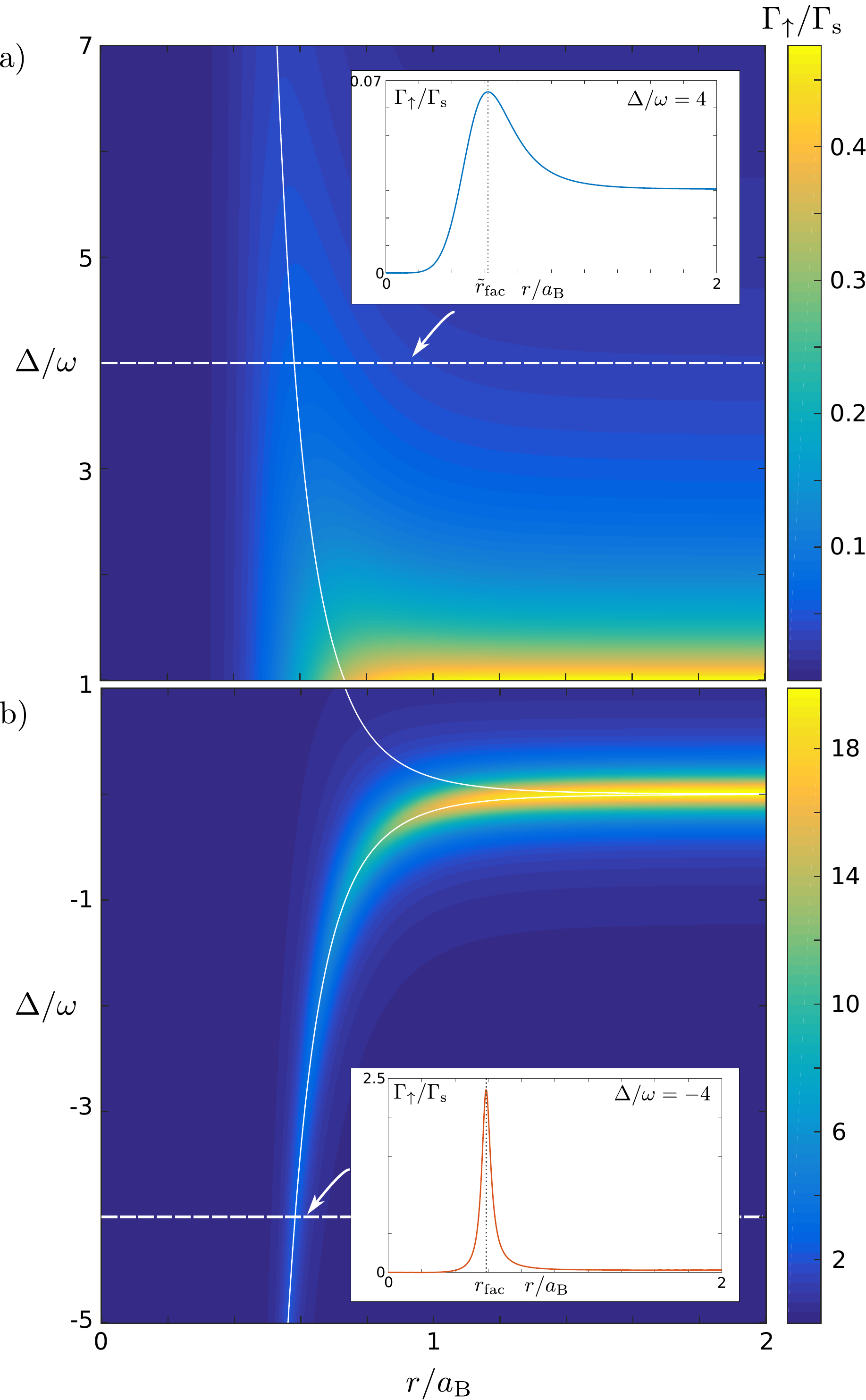, width=.45\textwidth}
\caption{ Incoherent excitation rate $\Gamma_\uparrow/\Gamma_\mathrm{s}(\Delta,r)$ in the presence of one Rydberg atom at distance $r$ with detuning $\Delta$ for attractive interaction $C_6 = - 10\omega$ with parameters as in Fig. \ref{fig:Fig1} and $D_6 = 0.3 C_6$. (a) \textit{Anomalous facilitation} on the blue detuned side and (b) \textit{usual facilitation} on the red detuned side with attractive interaction. The insets show the excitation profile $\Gamma_\uparrow(r)$ for a) blue ($\Delta/\omega = -4$) and b) red ($\Delta/\omega = 4$) detuning. The white solid line follows $|C_6/\Delta|^{(1/6)}$. }
\label{fig:Fig2}
\end{figure}

\begin{figure*}[t]
\centering
\epsfig{file=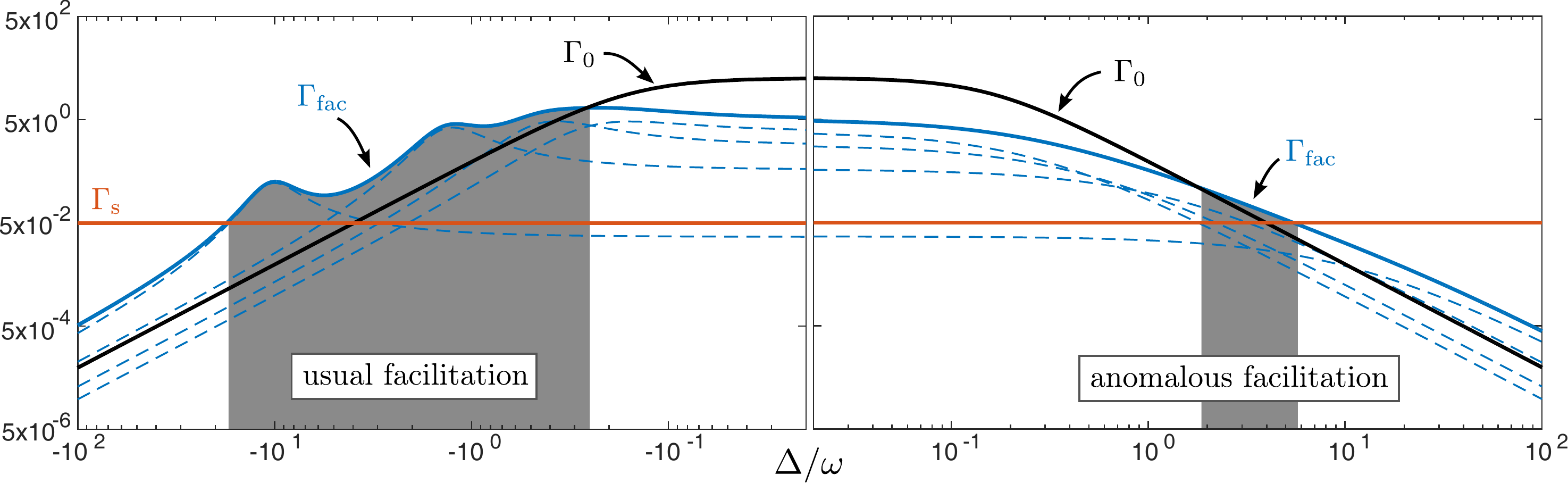, width=.95\textwidth}
\caption{ For the same parameters as in Fig. \ref{fig:Fig2} and a cubic lattice with lattice spacing $a$, we determine the total facilitation rate $\Gamma_\mathrm{fac}$  for all atoms within a blockade radius $a_\mathrm{B}/a \simeq 2$. We determine the region (gray shaded) where both conditions for facilitation (i) $\Gamma_\mathrm{fac} > \Gamma_0$ and (ii) $\Gamma_\mathrm{fac} > \Gamma_\mathrm{s}$ are fulfilled. The blue dashed lines show the facilitation rate for different distances $r_j$ to the excited Rydberg atom. Due to the lattice structure, we encounter a discrete set of resonances. The blue solid line is the total facilitation rate. }
\label{fig:Fig3}
\end{figure*}

\section{Inhomogeneous Broadening}

In the following, we discuss an additional inhomogeneous broadening motivated by interaction induced processes observed in several experiments \cite{Singer2004, Raitzsch2009, Goldschmidt2016}. The inhomogeneous broadening strongly depends on the microscopic details of the interaction potential. Therefore, a full theory of possible broadening mechanisms in Rydberg gas experiments is still missing. Yet, there are several possible scenarios: On the one hand, the strong and long-range interaction may induce additional dephasing mechanisms \cite{Raitzsch2009, Singer2004}. On the other hand, black-body induced or spontaneous transitions to neighboring Rydberg levels may lead to a dipole-dipole interaction induced broadening \cite{Goldschmidt2016}. Another possibility is the effect of motional decoherence in optical lattices with finite width \cite{Li2013, Macri2014,Letscher2016}. Crucially, they all originate from a strong and quasi long range interaction potential $V(r) = C_\alpha/r^\alpha$ and therefore depend on the mutual distance $r$ between two atoms.

Here, let us consider a special type of interaction induced broadening. We assume a decoherence rate
\begin{equation}
\label{eq:AdditionalBroadening}
\gamma(r) = \gamma_0 + \frac{D_\beta}{r^\beta}.
\end{equation}
Besides the bare decoherence rate $\gamma_0$, we added a dephasing rate seen by the ground state atom with distance $r$ to the excited atom. The coefficient $\beta$ determines the range and $D_\beta$ the strength of the broadening. Specifically, we discuss here the case $\beta = 6$ assuming a van der Waals type interaction. However, similar results can be obtained for different exponents $\beta$. Importantly, with decreasing distance between two atoms, the interaction induced broadening $\gamma(r)$ increases strongly. In this case, the full excitation rate is 
\begin{equation}
\label{eq:ExcitationRateBroadening}
\Gamma_\uparrow(r) = \frac{2\Omega^2\gamma(r)}{\gamma(r)^2+(\Delta - V(r))^2}.
\end{equation}
Note, we assume a symmetric inhomogeneous broadening mechanism centered at the resonance position $\Delta = V(r_\mathrm{fac})$. In the limit $r \rightarrow \infty$, we retrieve the excitation rate for uncorrelated atoms.

\section{Two \& Many Particle Dynamics: Anomalous Case}

In the following, we will show that attractive interactions with strong inhomogeneous broadening can result in an anomalous facilitation on the blue detuned side of the resonance. In particular, we discuss an attractive interaction potential $V(r) = C_6/r^6 < 0$ and additional inhomogeneous decoherence rate $\gamma(r)$ as in eqn. \eqref{eq:AdditionalBroadening}. Now, the line broadening $\gamma(r)$ strongly increases with decreasing mutual distance $r$ between two atoms in the ground and Rydberg state. The resulting excitation rate \eqref{eq:ExcitationRateBroadening} is shown in Fig. \ref{fig:Fig2}. 

On the red detuned side ($\Delta < 0$), see Fig. \ref{fig:Fig2}b, we recognize the usual facilitation mechanism. Here, the broadening mainly reduces the excitation rate compared to the excitation rate \eqref{eq:eq1} without additional broadening. At the facilitation radius $r_\mathrm{fac}$, the excitation rate $\Gamma_\uparrow(r)$ reaches approximately its maximum. However, we obtain a broader spatial width $\delta r \simeq r_\mathrm{fac} \gamma(r_\mathrm{fac})/ 6\Delta_0$ in linear order in $\delta r/r_\mathrm{fac}$.

Interestingly, we can also identify an anomalous facilitation radius $\tilde{r}_\mathrm{fac}$ on the blue detuned side $\Delta > 0$, as shown in the inset of Fig. \ref{fig:Fig2}a. Similar to the usual case, the anomalous facilitation at $\tilde{r}_\mathrm{fac}$ decreases with increasing detuning $\Delta$ approximately following the same scaling as $r_\mathrm{fac}$. In contrast to the usual facilitation, which results from energetic conditions, the anomalous radius $\tilde{r}_\mathrm{fac}$ originates from the inhomogeneity of the broadening mechanism. Overall, the excitation rate for an atom at distance $\tilde{r}_\mathrm{fac}$ with blue detuning is of course much smaller than in the red detuned case. However, the spatial width of the facilitation profile $\Gamma_\uparrow(r)$ is broadened, too. Naively, we do not expect any cascaded excitations since the rates are strongly reduced on the blue detuned side of the resonance for attractive interaction. However, we will show that facilitation can be possible for certain detunings $\Delta$ in sufficiently dense atomic gases.

So far, we discussed the excitation rate of a single ground state atom in the presence of one other already excited Rydberg atom at distance $r$. However, a single Rydberg excitation influences the excitation rate of \emph{all} surrounding atoms within a blockade radius $a_\mathrm{B}$. Therefore, we have to consider the integrated facilitation rate 
\begin{equation}
\Gamma_\mathrm{fac} = \sum_j^N \Gamma_\uparrow(r_j)
\end{equation}
over all atoms $j$ with distance $r_j$ to the excited Rydberg atom. Here $N$ is the number of atoms within the blockade radius $a_\mathrm{B}$. Atoms with distance $r > a_\mathrm{B}$, i.e. larger than the blockade radius, are only weakly affected by the interaction shift and therefore neglected. To observe facilitated excitations, we require two ingredients: Firstly, for an excitation cascade, we have to compare the integrated facilitation rate $\Gamma_\mathrm{fac}$ to the uncorrelated total excitation rate $\Gamma_0 = N\Gamma_\uparrow(r \rightarrow \infty)$. Secondly, we have to ensure, that one Rydberg excitation triggers a subsequent excitation before it decays with rate $\Gamma_\mathrm{s}$ back to the ground state. This leads to the following two conditions for facilitation
\begin{align}
\label{eq:condition1}
\mathrm{(i)}  \  \Gamma_\mathrm{fac} > \Gamma_0, \\ 
\label{eq:condition2}
\mathrm{(ii)} \  \Gamma_\mathrm{fac} > \Gamma_\mathrm{s}.
\end{align}

The conditions \eqref{eq:condition1} and \eqref{eq:condition2} may be fulfilled easily in the usual facilitation case. In the anomalous case, the individual excitation rates are much weaker. Nevertheless, since the spatial excitation width of the anomalous case is much broader, we may fulfill conditions \eqref{eq:condition1} and \eqref{eq:condition2} for a sufficiently dense atomic gas. 

To give an example, we calculate the integrated excitation rates $\Gamma_\mathrm{fac}$ and $\Gamma_0$ for a cubic lattice with lattice spacing $a_\mathrm{B}/a \simeq 2$ using realistic experimental parameters with weight $D_6/C_6 = 0.3$. Within a blockade radius $a_\mathrm{B}$, we count $N=32$ atoms. The results are shown for both, blue and red detuning, in Fig. \ref{fig:Fig3}. While we can identify individual resonance peaks (dashed blue lines) on the red detuned side corresponding to specific lattice distances, on the blue detuned side only the residual line broadening of the resonances remain. Due to the inhomogeneous broadening, the resonance lines on the red detuned side are strongly broadened. Here, conditions eqn. \eqref{eq:condition1} and \eqref{eq:condition2} are fulfilled for a broad range of detunings (gray shaded region) and we observe usual faciliation. Surprisingly, on the blue detuned side both facilitation conditions in eqn. \eqref{eq:condition1} and \eqref{eq:condition2} are fulfilled for several detunings, too. Clearly, the extent of allowed detunings $\Delta$ is much larger on the red detuned side. However, an anomalous facilitation is possible for a finite region of detunings $\Delta$. Note that the anomalous facilitation range strongly depends on the density, the microscopic geometry, the interaction and the line broadening.

\section{Summary \& Outlook}

To conclude, we discovered an anomalous facilitation mechanism allowing to observe cascaded excitations in an experimentally accessible parameter regime. Our discussion was motivated by the observation of an interaction induced inhomogeneous broadening mechanism in a recent experiment \cite{Goldschmidt2016}. While we here assumed an additional decoherence proportional to the van der Waals interaction, we would like to point out that other spatially dependent broadening mechanisms may lead to similar results. We discussed the general conditions for facilitation in the many particle dynamics.

We believe that the anomalous facilitation is an interesting extension to the usual facilitation, which may be observable in current experiments. In particular, it may lead to unexpected correlations on the "wrong" side of the resonance. For instance, attractive interactions may lead to the formation of molecular bound states suppressing an excitation cascade on the red detuned side. However, the observation of strongly correlated excitation growth may be seen on the blue detuned side of the resonance. Finally, we hope that our discussion further motivates to study the origin of excitation line broadening in ultra cold Rydberg gases and may trigger applications thereof. 

\section{Acknowledgments}
We acknowledge financial support by the DFG within the SFB/TR49 and SFB/TR185. The authors thank C. Lippe and D. Petrosyan for helpful discussions. F.L. and O.T. are recipients of a fellowhship through the Excellence Initiative MAINZ (DFG/GSC 266). 

\bibliography{Broadening.bib}

\end{document}